\documentclass{PoS}

\usepackage{cite}
\usepackage{amsmath,amssymb}

\newcommand{\be}{\begin{equation}}
\newcommand{\ee}{\end{equation}}
\newcommand{\ba}{\begin{eqnarray}}
\newcommand{\ea}{\end{eqnarray}}

\def\lsi{\raise0.3ex\hbox{$<$\kern-0.75em\raise-1.1ex\hbox{$\sim$}}}
\def\gsi{\raise0.3ex\hbox{$>$\kern-0.75em\raise-1.1ex\hbox{$\sim$}}}
\newcommand{\lsim}{\mathop{\lsi}}

\title{
{\vspace{-25mm} \normalsize\hfill{\small CERN-PH-TH/2011-274}}\\[20mm]
Towards corrections to the strong coupling limit of staggered lattice QCD}

\ShortTitle{Corrections to QCD strong coupling limit}

\author{Michael Fromm, Jens Langelage, \speaker{Owe Philipsen}\\
        Institut f\"ur Theoretische Physik, Johann Wolfgang Goethe-Universit\"at Frankfurt,\\
        60438 Frankfurt am Main, Germany\\
        E-mail: \email{fromm,langelage,philipsen@th.physik.uni-frankfurt.de}}

\author{Philippe de Forcrand, Wolfgang Unger\\
        Institute for Theoretical Physics, ETH Z\"urich, CH-8093 Z\"urich, Switzerland\\
        Physics Department, TH Unit, CERN, CH-1211 Geneva 23, Switzerland\\
        E-mail: \email{forcrand,ungerw@phys.ethz.ch}}

\author{Kohtaroh Miura\\ INFN Laboratori Nazionali di Frascati, I-00044 Frascati (RM), Italy\\
          E-mail: \email{kohtaroh.miura@lnf.infn.it}}

\FullConference{The XXVIII International Symposium on Lattice Field Theory, Lattice2010\\
		June 14-19, 2010\\
		Villasimius, Italy}

\abstract{
We report on the first steps of an ongoing project to add gauge observables and gauge 
corrections to the well-studied strong coupling limit of staggered lattice QCD, which has 
been shown earlier to be amenable to numerical simulations by the worm algorithm
in the chiral limit and at finite density. 
Here we show how to evaluate the expectation value
of the Polyakov loop in the framework of the strong coupling limit at finite temperature, 
allowing to study confinement properties along with those of chiral symmetry breaking. We find 
the Polyakov loop to rise smoothly, thus signalling deconfinement. The non-analytic nature
of the chiral phase transition is reflected in the derivative of the Polyakov loop.
We also discuss how to construct an effective theory for non-zero lattice coupling, which is 
valid to $O(\beta)$. 
}

\FullConference{ The XXIX International Symposium on Lattice Field Theory - Lattice 2011\\
July 10-16, 2011\\
Squaw Valley, Lake Tahoe, California}

\begin{document}
\section{Introduction}

Due to the sign problem prohibiting lattice simulations at finite baryon density, 
the QCD phase diagram in the space of temperature $T$ and chemical potential for baryon number
$\mu_B$ remains largely unknown. 
Employing indirect methods like reweighting, Taylor expansions about
$\mu_B=0$ or simulations at imaginary chemical potentials $\mu=i\mu_i,\mu_i\in\mathbb{R}$, followed by 
analytic continuation, controlled calculations are only feasible as long as the quark chemical potential
$\mu=\mu_B/3\lsim T$ \cite{review1,review2}. 
The latter two methods predict an at least initial weakening of the QCD quark hadron transition for quark chemical potentials $\mu\lsim T$ \cite{fp3,fp4,endrodi}. 
A possible critical point at larger chemical potentials as well as a phase with net baryon number 
at low temperatures remain inaccessible to those methods. 
Furthermore, the chiral limit is inaccessible to direct simulations, thus obscuring its influence on the chiral
properties of the QCD phase transition. 
These problems motivate the continued study of
effective theories which are more amenable to finite density studies than full QCD.

In order to control systematic errors, one would like to derive such effective theories
directly from QCD, ideally with the possibility of systematic improvements  by iteration of the 
approximation scheme. Such a scheme is afforded by the strong coupling expansion. Recently,
it has been used to derive a centre-symmetric 3d effective action for Yang-Mills theory, which 
provides a quantitative description of the pure gauge deconfinement transition to within
10\% of the corresponding full 4d simulations \cite{zn}. The extension to include
Wilson fermions has also been reported at this conference \cite{kappa}, but is limited to heavy fermions.

In this contribution we consider lattice QCD with staggered fermions
in the strong coupling limit, i.e.~for plaquette coupling $\beta=0$. This is lattice QCD with
one parameter chosen far away from its physical value. This is well motivated since the resulting model
is confining, its massless limit exhibits chiral symmetry as well as its spontaneous breaking and, most importantly,  it may be simulated at finite baryon densities by a suitable choice of algorithm.

Starting from this model we can take into account gauge corrections, firstly with the aim of making contact with the successful description of the Yang-Mills action in terms of
a strong coupling series \cite{zn}, and secondly to gain insight into worm-inspired algorithmic methods
\cite{worm} for Lattice QCD involving both, a fermionic and a gauge part in its action. Inclusion of fermionic
effects and gauge corrections have also been pursued analytically in the mean field approximation\cite{mf}. Here, we aim to extend these investigations to a full evaluation of the effective theory.

\begin{figure}[t!!!]
\vspace*{-0.7cm}
\centerline{
\includegraphics[width=0.5\textwidth]{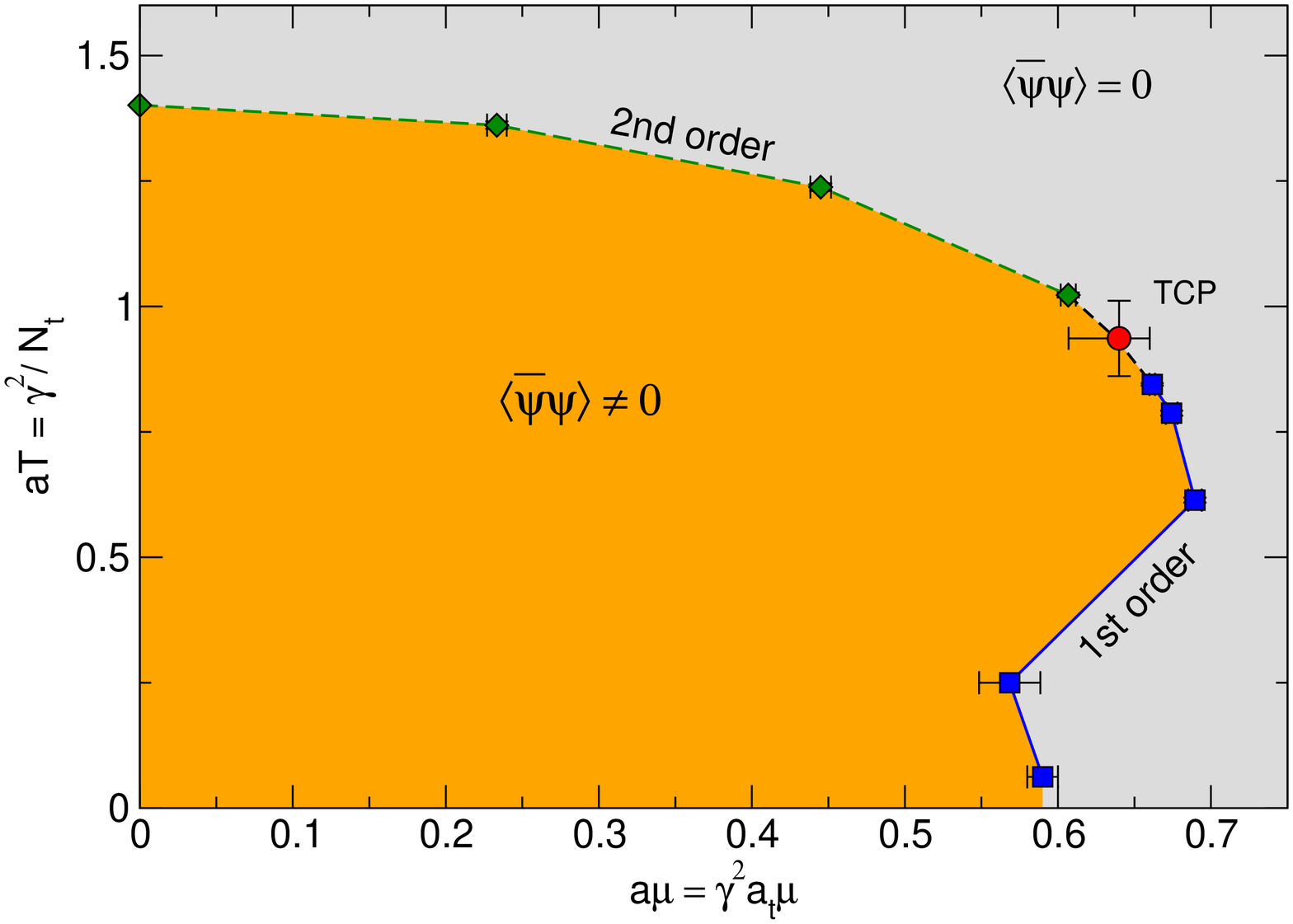}
\includegraphics[width=0.5\textwidth]{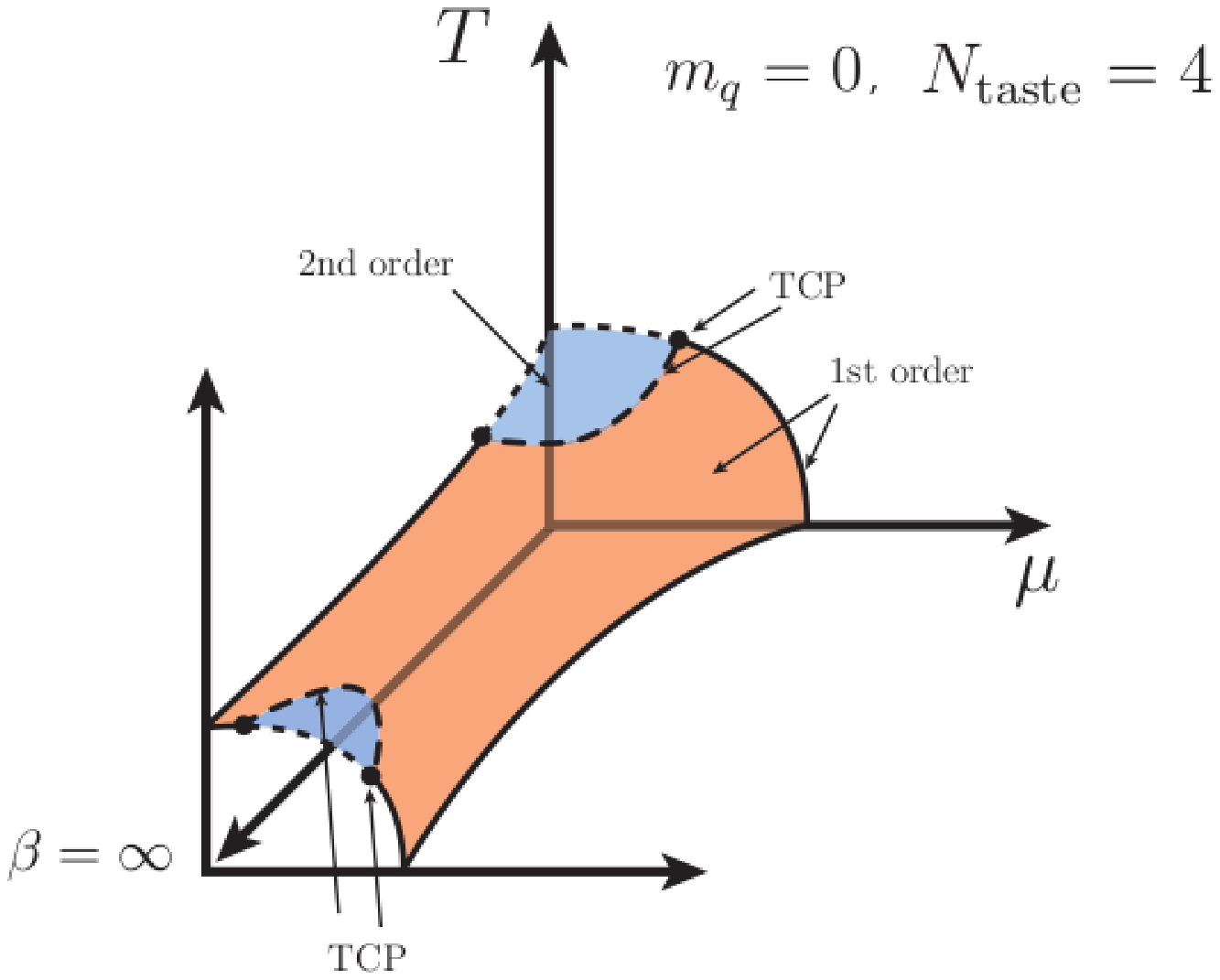}
}
\caption[]{Left: The phase diagram for $N_f=1$ QCD in the strong coupling and chiral limit  on an
$N_t=4$ lattice \cite{ff}.
Right: One possible extension for finite values of $\beta$. For large $\beta$ the phase diagram connects to that
of the $N_f=4$ theory in the continuum limit.
}
\label{pd}
\end{figure}

Starting point is lattice QCD with staggered quarks and the Wilson action, $S=S_G+S_F$,
\ba
S_G&=&-\frac{\beta}{3}\sum_p {\rm Re Tr} U_p\\
S_F&=&\sum_x\left\{\sum_\mu \eta_\mu(x)[\bar{\chi}(x)U_\mu(x)\chi(x+\hat{\mu})
-\bar{\chi}(x+\hat{\mu})U_\mu^\dag(x)\chi(x)]+2am\bar{\chi}(x)\chi(x)\right\}.
\ea
Expanding the partition function in a power series of $\beta=6/g^2$,
\be
Z=\int D\chi D\bar{\chi} DU\,(1-S_G+O(\beta^2))e^{-S_F},
\label{betaseries}
\ee
the leading order expression corresponding to the strong coupling limit, $\beta=0$, is very simple
due to the neglect of the gauge action. In this case there are only one-link integrals over the coupling
terms in the fermion action, which can be done exactly to produce a purely fermionic theory,
\be
Z(\beta=0)=\int D\chi D\bar{\chi} DU\,e^{-S_F}=\int D\chi D\bar{\chi} \prod_{x,\nu} \int dU_\nu(x)\,e^{-S_F}
=\int D\chi D\bar{\chi}\prod_{x,\nu} z(x,\nu)\;.\label{eq:Zsc}
\ee
Its integrand now depends on colourless hadronic degrees of freedom only, mesons $M_x=\bar{\chi}\chi(x)$ and
baryons $B_x=\frac{1}{3!}\epsilon_{abc}\chi_a\chi_b\chi_c(x)$,
\be
z(x,y)=\sum_{i=0}^3 \alpha_i(M_x,M_y)^i+\tilde{\alpha}\bar{B}_xB_y+\tilde{\beta}\bar{B}_yB_x\;.
\ee
The final Grassmann integration can be done introducing a world line formulation
to yield
\be
Z(\beta=0)=\sum_{\{C\}}\prod_{b=(x,\nu)}w_b\prod_x w_x \prod_l w_l,
\label{eq:Zsc1}
\ee
with $w_x, w_b, w_l$ contributions from the chiral condensate, meson hopping and baryon world lines, respectively \cite{rossi}. 
This so-called monomer-dimer model, representing the strong coupling limit of QCD,
displays confinement as well as, in the massless limit for one flavour $N_f=1$, a $U(1)$ chiral symmetry
which  spontaneously breaks in the vacuum. It is thus an interesting theory to investigate 
the finite temperature restoration of chiral symmetry.
Since $\beta=0$, the definition of a temperature scale requires introduction of an anisotropic lattice.
The anisotropy parameter is defined in the weak coupling
limit, $\gamma=a/a_t$. At strong coupling and in the mean field approximation one has instead
$\gamma^2\approx a/a_t$ and hence $aT=\gamma^2/N_t$. The model in terms of baryonic degrees
of freedom has a sign problem even at $\mu=0$, which however can be solved by an 
analytic reordering of terms \cite{km}.
For $\mu\neq0$, there is again a mild sign problem. Using a worm algorithm \cite{worm} allows for simulations of large lattices (spatial volume $16^3$) at the relevant temperatures and densities~\cite{ff}. Alternatively, the model Eq.(\ref{eq:Zsc}) can be formulated and studied in continuous Euclidean time~\cite{unger}, taking the joint limit $\gamma, N_t\rightarrow \infty$, $aT=\gamma^2/N_t$ fixed. There baryons become static and the sign problem vanishes. The strength of the sign problem at finite $N_t$ can be parametrised by 
quoting the average sign measured in the phase quenched theory, which can be expressed
as a ratio of partition functions,
\be
\langle \mbox{sign}\rangle_{||}=\frac{Z}{Z_{||}}=e^{-\frac{V}{T}\Delta f(\mu^2)}, \quad Z_{||}: \mbox{phase quenched},
\quad \Delta f\sim 10^{-4}.
\ee
Using this approach, the phase diagram of the $N_f=1$ theory in the chiral limit has been mapped 
out \cite{ff}, revealing a second order chiral phase transition for small $\mu$, which turns into a first order
phase transition at a tricritical point, Fig.\ref{pd}. Note that, since there is the breaking/restoration of a symmetry involved, there is a true non-analytical phase transition separating the symmetric and broken 
phases. If one switches on small quark masses, chiral symmetry is broken explicitly and the 
second order phase transition changes into an analytical crossover.

The long term goal of the present study is to extend the model Eq.~(\ref{eq:Zsc}) to finite values of the gauge
couplings by including correction terms in the effective action, and possibly constrain the phase
diagram in the weak coupling limit, as shown schematically in Fig.\ref{pd} (right).

\section{Gauge observables in the strong coupling limit}

\begin{figure}[t!!!]
\vspace*{-0.7cm}
\centerline{
\includegraphics[width=0.5\textwidth]{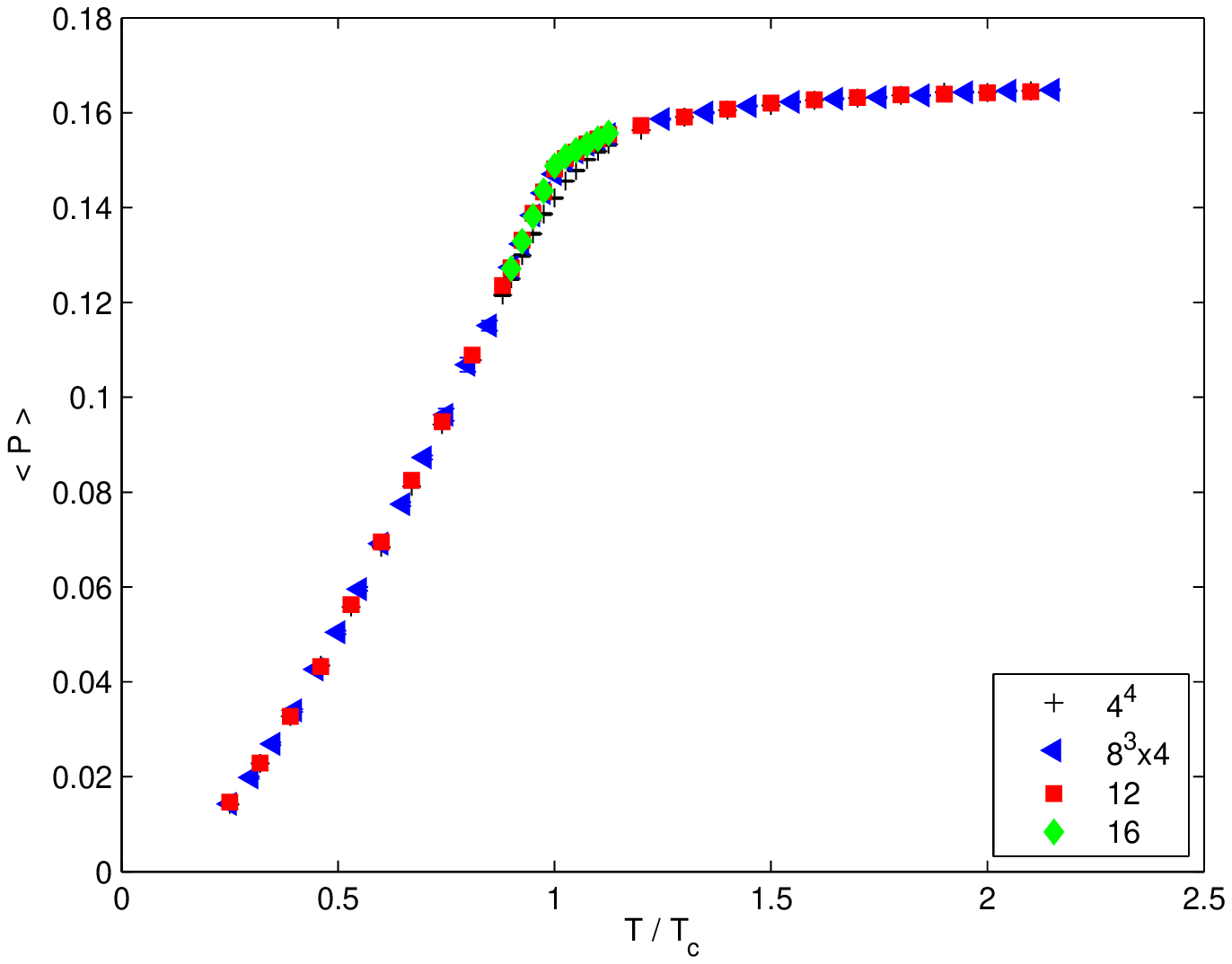}
\includegraphics[width=0.5\textwidth]{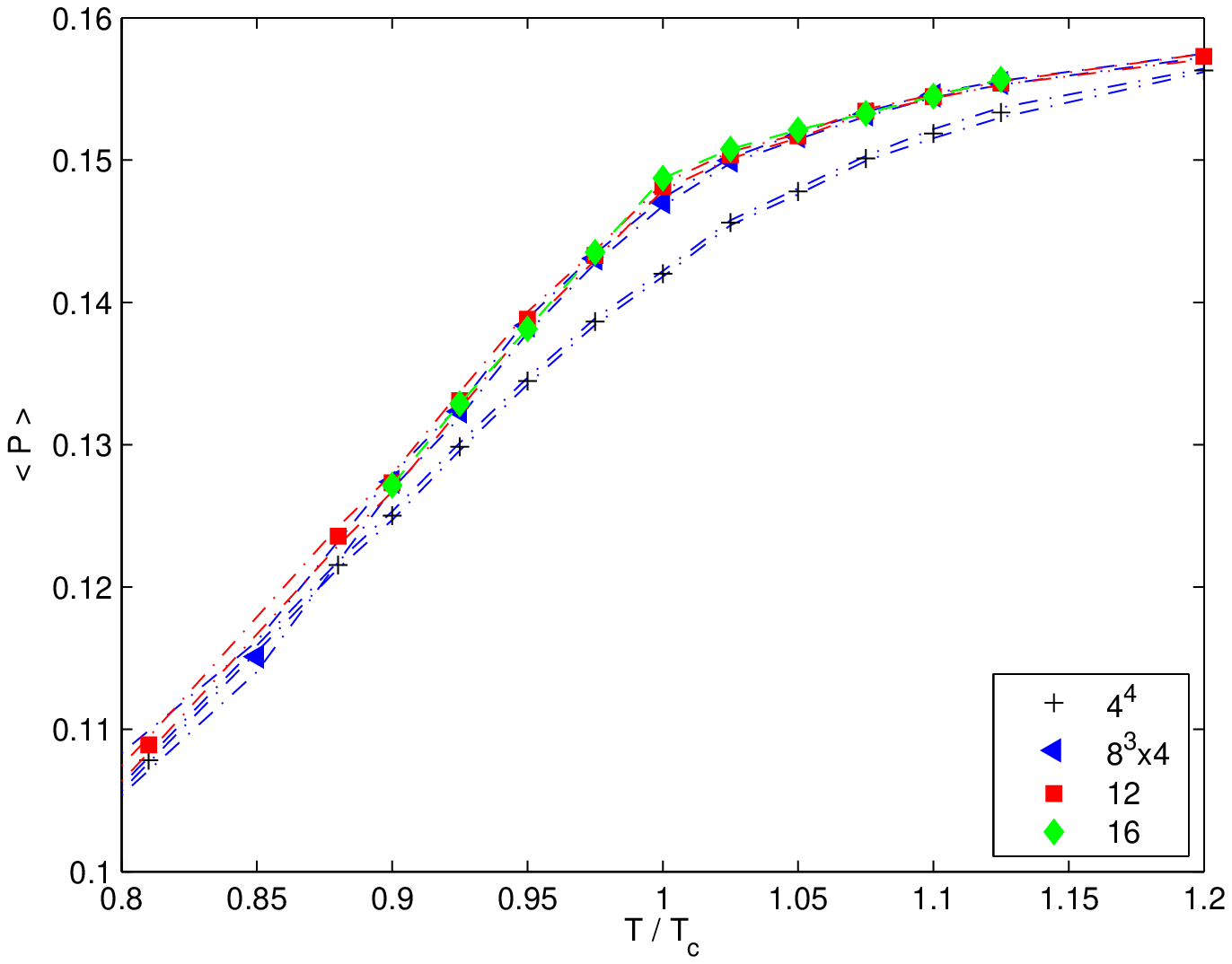}
}
\caption[]{Left: The Polyakov  loop $\langle P \rangle$ as a function of temperature across the chiral transition for several volumes ($am = 0$). Right: Close-up of the transition region.}
\label{poly}
\end{figure}

\begin{figure}[t!!!]
\vspace*{-0.3cm}
\centerline{
\includegraphics[width=0.5\textwidth]{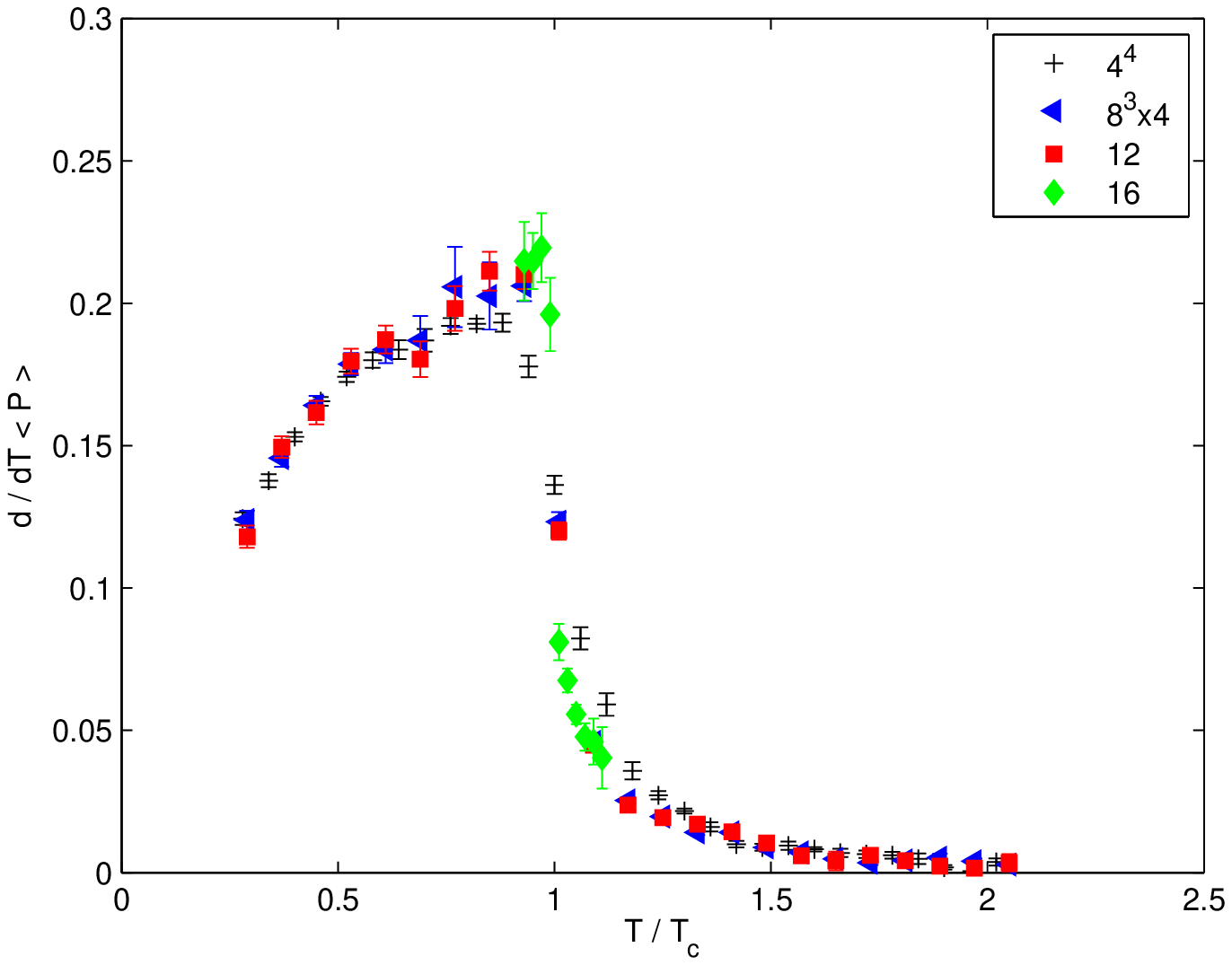}
\includegraphics[width=0.5\textwidth]{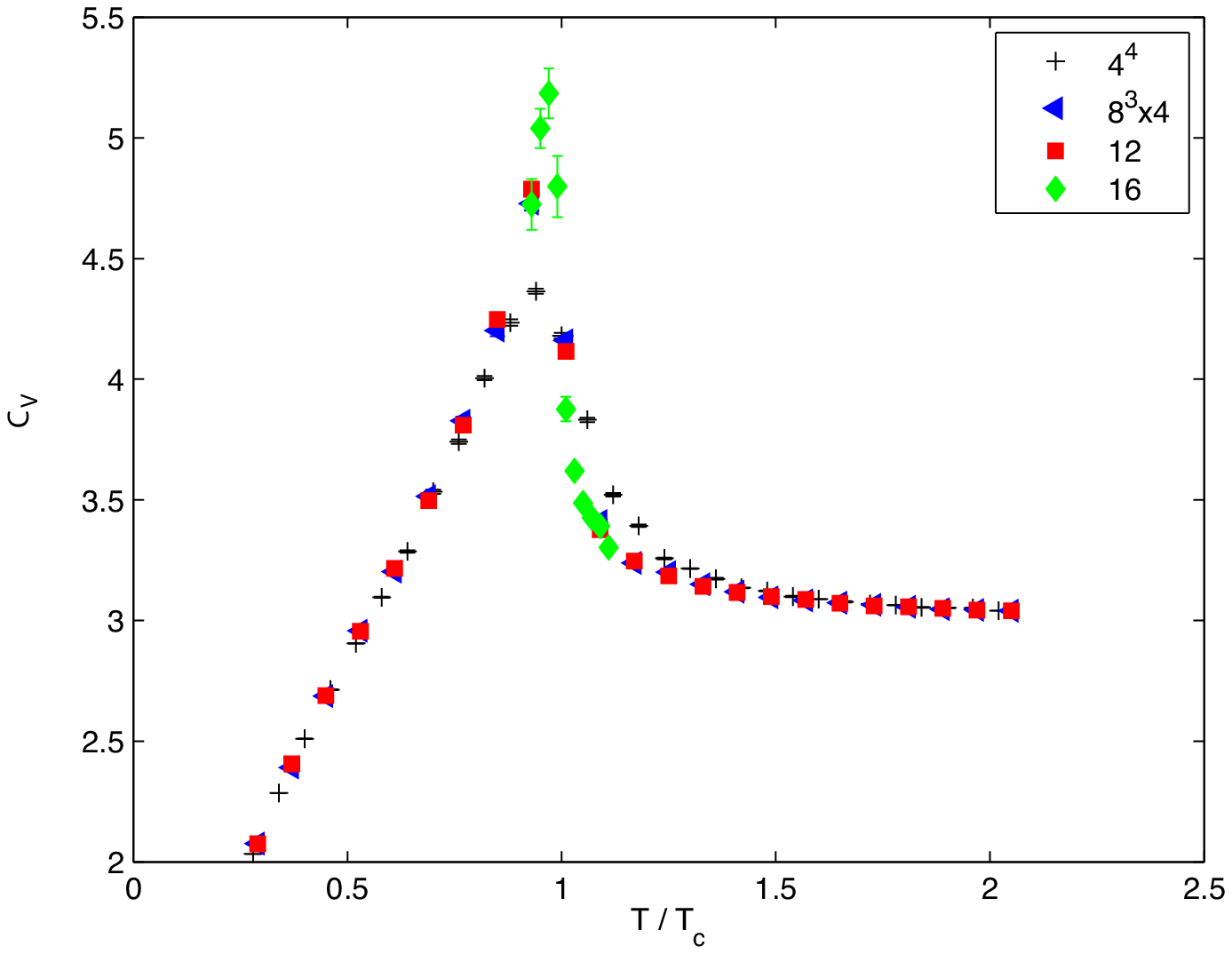}
}
\caption[]{Left: The derivative $\mathrm{d} \langle P\rangle/\mathrm{d}T $ across the chiral transition. Right: The derivative of the internal energy $\mathrm{d}\epsilon/\mathrm{d}T $ shows a divergence consistent with $\alpha \approx 0$.}
\label{dP}
\end{figure}

As a preparatory study towards including gauge corrections, we investigate the possibility of
describing gauge field dynamics in the effective theory. In particular, one would like to study 
the confining behaviour of the theory across the chiral transition. However, in order to evaluate gauge observables in the strong coupling limit, we have to go back and write
down their expectation values using the partition function in the strong coupling limit, but {\it before} 
gauge integration.
Next, we rewrite the expression in terms of $Z_F$, corresponding to the partition function in the strong coupling limit before fermion integration,
\ba
\langle O \rangle &=&\frac{1}{Z}\int D\bar{\chi}\,D\chi\int DU \, O[U]\, e^{-S_F[U,\chi,\bar{\chi}]}
=\frac{\int D\bar{\chi}\,D\chi \; \langle O\rangle_U Z_F}{\int D\bar{\chi}\,D\chi\;Z_F} ,\\
\langle O \rangle_U&=&\frac{1}{Z_F}\int DU\, O[U]\,e^{-S_F},\quad
Z_F=\int DU\, e^{-S_F}=\prod_{x,\nu}z(x,\nu)\;.
\ea
That is, for every gauge observable we wish to compute, the gauge integration has to be done
with a different and now non-trivial integrand.
Let us illustrate this with the Polyakov loop. We use a generating functional to define
the expectation value by a functional derivative,
\ba
\langle P \rangle&=&\frac{d}{dJ}\ln Z_J\Big|_{J=0}\\
Z_J&=&\int D\bar{\chi}\,D\chi\int DU\, e^{-S_F[U,\chi,\bar{\chi}]+JP^\dag+J^\dag P}
\approx \int D\bar{\chi}\,D\chi \, Z_F(\langle JP^\dag\rangle_U+\langle J^\dag P\rangle_U),
\label{Ploop}
\ea
where we have approximated the exponential by its leading term, as we only consider infinitesimal
$J$. After Grassmann-integration $Z_J$ takes on the form of Eq.(\ref{eq:Zsc1}), this time with modified, $J$-dependent weights $\tilde{w}_b$, $\tilde{w}_x$, $\tilde{w}_l$, and can again be evaluated using worm methods. Fig.\ref{poly} shows the
behaviour of $\langle P \rangle$ across the chiral phase transition as the temperature is varied. Interestingly,
the Polyakov loop rises significantly, thus indicating deconfinement, even though we are in the 
strong coupling limit. However, the behaviour through the phase transition is smooth, similar
to the findings by renormalisation group methods \cite{paw}. However, looking at the derivative of $\langle P\rangle$ in Fig.\ref{dP} (left), 
we note that it develops a cusp. Thus, also the 
Polyakov loop signals non-analytic behaviour through the chiral phase transition, albeit in a higher
derivative than the order parameter does. 
This can be 
understood by noting that the Polyakov loop can be written as a ratio of partition functions with and 
without a static quark $Q$,
\be
\langle P \rangle\sim \exp -(F_Q-F_0)/T=\frac{Z_Q}{Z}\;.
\ee
Since the free energy and its first derivative are continuous 
through a second order phase transition, so is the Polyakov loop.
For comparison we also study the average number of meson hops in $t$-direction which is a measure for the internal energy $\epsilon$ of the model. Its derivative 
$C_V=\mathrm{d}\epsilon/\mathrm{d}T$, shown in Fig.~\ref{dP} (right) as a function of temperature, is peaked at the transition with critical exponent $\alpha\approx 0$ as expected in the 3d $O(2)$-universality class.

\section{Computing leading order gauge corrections}

\begin{figure}[t!!!]
\vspace*{-0.7cm}
\includegraphics[height=0.2\textwidth]{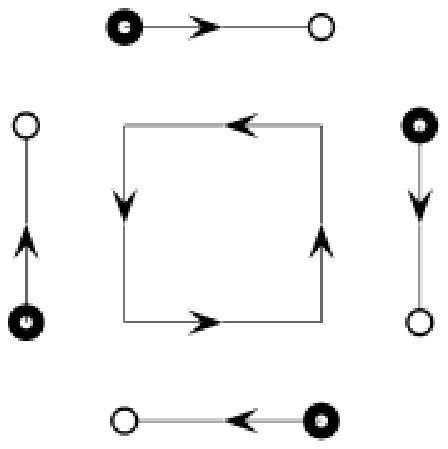}
\includegraphics[height=0.2\textwidth]{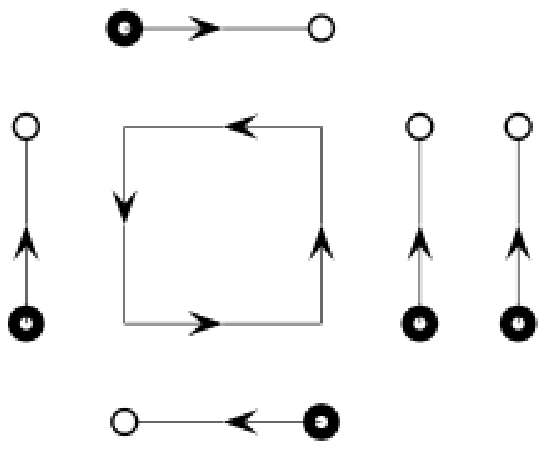}\hspace*{1cm}
\includegraphics[width=0.7\textwidth]{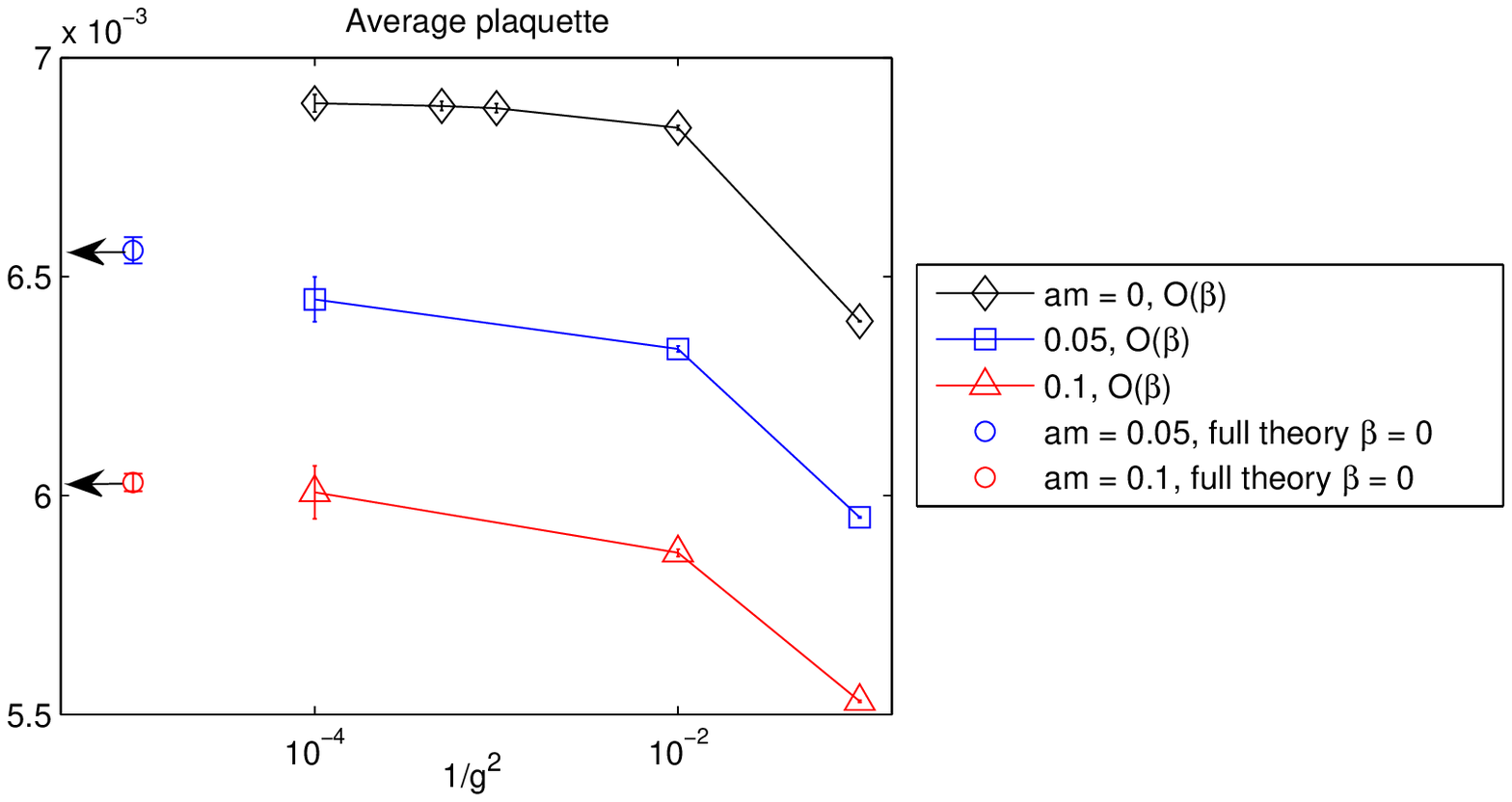}
\caption[]{Left: Two $\mathcal{O}(\beta)$-contributions to the effective action. Quark hops of the Dirac-part (arrows connecting circles) pair up with a plaquette-term to form colour-singlet degrees of freedom. Right: Average plaquette as a function of $1/g^2$ using the $\mathcal{O}(\beta)$-contribution of the $U(3)$ (purely mesonic) theory on a $8^3\times16$-lattice. For $\beta\rightarrow 0$ they agree with the average plaquette obtained for $SU(3)$ using HMC~\cite{kim}.} 
\label{num}
\end{figure}

We are now ready to address the calculation of gauge corrections to the strong coupling limit.
To this end, we return to our initial expression, Eq.~(\ref{betaseries}). For gauge observables in the strong coupling limit it was sufficient to treat the relevant observables as source terms, c.f.~the Polyakov loop, Eq.~(\ref{Ploop}), expand the relevant term to first order in $J$ and take the limit $J\rightarrow 0$ numerically. Here, we seek to take into account $\mathcal{O}(\beta)$-corrections to the strong coupling action. We must hence study
\be
Z=\int D\chi D\bar{\chi}Z_F\langle\exp{(-S_G)}\rangle_U \approx \int D\chi D\bar{\chi}\,Z _F \exp{\left(\langle -S_G\rangle_U\right)}\;,
\ee
where in the approximation we used $\langle\exp{(-S_G)}\rangle_U = \exp{\left( \sum_{k=1}^\infty \frac{1}{k!}\langle (-S_G)^k\rangle_{U,c}\right)}$ and truncated at $k=1$, i.e. to order $1/g^2$.
The implementation of gauge corrections now consists of two parts. First, we have to evaluate
the gauge integrations present in the term $\langle (-S_G)^k\rangle_{U,c}$ which are non-trivial.
Introducing the plaquette term leads to 19 new diagrams of the type shown in Fig.\ref{num} (left).
After the gauge integration, they enter as new terms to the effective action.
Note that this leading order correction has profound consequences on the 
physics of the effective theory. The gluon loop now allows for a separation of coloured quarks by one lattice spacing $a$. In particular, baryon hoppings on the lattice 
are now no longer self-avoiding, thus lending spatial structure to the baryon.

The second part consists of the numerical evaluation of the partition function (or observables), now
making use of the enlarged effective action. This is a non-trivial task, as it is not clear whether
all additional terms can be regrouped in terms of the world line formulation that is then amenable
to simulations by the worm algorithm. In our first attempt at a numerical evaluation, we have so far
implemented only the mesonic contributions to the partition function, neglecting the baryons.
In certain regimes of interest, in particular at zero net baryon density, 
this is a good approximation because baryons are heavy compared to mesons and therefore
influence the dynamics only weakly.

In Fig.~\ref{num} we show our numerical results which, in the case of small but non-vanishing quark masses, we compare with those of full $SU(3)$
HMC simulations, represented by the data points on the left. We see that
for $\beta\rightarrow 0$ we obtain for the expectation value of the plaquette $\langle U_p\rangle_U\rightarrow \langle U_p\rangle_{HMC}$, as expected. This is a quite non-trivial check, as these agreeing
expectation values are simulated in terms of completely different degrees of freedom. The next steps to
be taken are to implement the baryonic contributions, which will enable us to investigate finite
coupling corrections to the phase diagram in the chiral limit. 

\section{Conclusions}

The strong coupling limit of QCD is interesting, because it represents a confining theory featuring
a chiral phase transition in the massless limit, which is amenable to simulation by the worm algorithm,
both at zero and finite baryon densities, even at low temperatures. We have shown that, even 
in the strong (i.e.~infinite bare) gauge coupling limit, the chiral phase transition also implies
a loss of confinement, as signalled by the rise of the Polyakov loop through the transition. 
Since the Polyakov loop is only indirectly associated with the order parameter of the chiral condensate,
its expectation value goes through the transition smoothly, while its derivative develops a cusp.

We are also on the way to including complete $O(\beta)$-corrections to the strong coupling limit, with
a large number of analytic plaquette type contributions to the effective theory. Simulations using the worm algorithm
including the mesonic part of the effective action give results consistent with full HMC simulations, 
and the implementation of the baryonic contributions is on the way.
From an analytical point of view, higher order corrections are feasible.
However, it remains to be seen whether the resulting additional terms to the effective action can 
be cast into a form amenable to be simulated by the worm algorithm, with a manageable sign problem. \\

\noindent
{\bf Acknowledgement:}  M.F.~,S.L.~and O.P.~are supported by the German BMBF, 06MS9150;
W.~U.~is supported by the Swiss Nat.~Fonds, 200020-122117.

\end{document}